\RequirePackage[2020-02-02]{latexrelease}
\documentclass[10pt]{iopart}
\usepackage{ucs}
\usepackage{graphicx}
\input epsf

\expandafter\let\csname equation*\endcsname\relax
\expandafter\let\csname endequation*\endcsname\relax
\usepackage{amsmath}
\usepackage{amssymb}
\usepackage{amsfonts}
\usepackage{float}
\usepackage{color,soul}
\usepackage{tabularx}

\begin{document}
\title[\rm{Supercond. Sci. Technol.}]{Superconductor to metal quantum phase transition with magnetic field in Josephson coupled lead islands on Graphene}

\author{Suraina Gupta, Santu Prasad Jana, Rukshana Pervin and Anjan K. Gupta}

\address{Department of Physics, Indian Institute of Technology Kanpur, Kanpur 208016, India}
\ead{anjankg@iitk.ac.in}
\vspace{10pt}
\begin{indented}
\item[] July 2024
\end{indented}

\begin{abstract}
Superconductor-to-metal transition with magnetic field and gate-voltage is studied in a Josephson junction array comprising of randomly distributed lead islands on exfoliated single-layer graphene with a back-gate. The low magnetic-field superconductivity onset temperature is fitted to the Werthamer-Helfand-Hohenberg theory to model the temperature dependence of the upper critical field. The magnetoresistance in the intermediate temperature and field regime is described using thermally activated flux flow dictated by field dependent activation barrier. The barrier also depends on the gate voltage which dictates the inter-island Josephson coupling and disorder. The magnetoresistance near the upper critical field at low temperatures shows signatures of a gate dependent continuous quantum phase transition between superconductor and metal. The finite size scaling analysis shows that this transition belongs to the $(2+1)$D-XY universality class without disorder.
\end{abstract}

\vspace{2pc}

\maketitle
\ioptwocol

\section{Introduction}
Quantum phase transitions involve change between two phases when a non-thermal parameter is varied at zero temperature \cite{sondhi1997continuous} through a critical value. Signatures of the quantum phase transition (QPT) can also be seen at finite but low temperatures. The finite size scaling analysis helps decipher the universality class of the QPT \cite{fisher1990quantum}. A variety of QPTs have been observed in two-dimensional (2D) superconductors (SCs) and Josephson junction (JJ) arrays that include superconductor-insulator transition (SIT) \cite{allain2012electrical,hen2021superconductor}, superconductor-metal transition (SMT) \cite{biscaras2013multiple,sun2018double} and Hall-insulator to SC transition \cite{breznay2016self}. The vortices and the duality between Cooper pair (CP) and vortex are pertinent to such transitions.

The dynamics of vortices is important for electrical transport in a 2D SC and thus it attracts interest for its intriguing physics \cite{blatter1994vortices,newrock2000two} and applications \cite{ma2020braiding,foltyn2007materials,wallraff2005quantum}. Vortices are topological excitations amounting to spatial phase-variation and these can arise due to magnetic fields, thermal or quantum fluctuations and electrical currents \cite{tinkham2004introduction}. Under an applied electrical current a vortex experiences force perpendicular to the current flow. Its resulting movement leads to time-variation of the local phase amounting to local electric field and thus electrical resistance. The zero-resistance state can still get restored if the vortex motion is checked due to either vortex pinning centers or inter-vortex interactions \cite{blatter1994vortices}.

The vortex states in 2D SCs are complex, involving ordered as well as disordered solid-like or fluid-like states \cite{blatter1994vortices}. The response of these vortices to bias current dictates the resistance of the SC. In a disordered 2D SC at low-field where vortex-density is small, the resistance is dictated by the dynamics of independent vortices in presence of pinning centers. With increasing field or vortex density, both the inter-vortex interaction and pinning become important. Eventually, near the upper critical field the CP de-pairing effects and quasi-particle physics will dominate as the SC order parameter diminishes. JJ arrays are ideal for studying this physics \cite{newrock2000two,fazio2001quantum} due to their tunable junction parameters giving control on effective phase-stiffness or on $E_{\rm J}/E_{\rm C}$, i.e. the ratio of the Josephson coupling energy to the Coulomb energy \cite{van1996quantum}. The same handle in homogeneous SC ultrathin films \cite{markovic1999superconductor,frydman2002universal} and granular SCs \cite{allain2012electrical,hen2021superconductor} is obtained by control of thickness, magnetic field or back gate voltage.

The `bosonic' picture of QPT attributes resistance in 2D SC with a non-vanishing SC order parameter to phase fluctuations \cite{fisher1990quantum}. This can result into an insulating state with delocalized vortices and localized CPs. At the SIT, a universal resistivity given by the quantum of resistance `$R_{\rm Q}$' \cite{fisher1990quantum,cha1994universal} is expected. A scaling theory based on interacting bosons by Fisher et al. \cite{fisher1990quantum,fisher1990presence} proposed a phase diagram for a 2D SC as a function of temperature, disorder, and magnetic field. Alternatively, the `fermionic' picture attributes such QPT to amplitude fluctuations, where CPs break into single electrons and the SC order parameter or the BCS energy gap vanishes \cite{valles1992electron,hsu1995magnetic,szabo2016fermionic}.

SITs in 2D SC systems such as tin-graphene hybrid \cite{allain2012electrical}, In-$\mathrm{InO_x}$ composite \cite{hen2021superconductor} and disordered $a$-$\mathrm{TaN_x}$ films \cite{breznay2017superconductor} belong to the universality class of $(2+1)$D XY model with disorder \cite{cha1994universal}. Here, the correlation length exponent $\nu \geq 1$ while $\nu < 1$ arises in systems without disorder, that is, in a clean regime \cite{cha1994universal}. Magnetic field-tuned phase transitions, such as SITs and SMTs, have also been observed in many systems \cite{hen2021superconductor,biscaras2013multiple,markovic1999superconductor,hao2020transport,jing2023quantum}. Easy regulation of charge carrier density in graphene through back-gate voltage was initially used by Allain et al. \cite{allain2012electrical} to study gate-tunable SIT in tin-graphene hybrid devices. Subsequent studies in similar systems also revealed double quantum criticality \cite{sun2018double} and quantum Griffith's singularity \cite{han2020disorder} under applied magnetic fields.

In this paper, electrical transport study as a function of magnetic field, temperature and back-gate voltage $V_{\rm g}$ is reported on a 2D device with lead (Pb) islands on single-layer graphene. The resistance at low field is understood using thermally activated de-pinning of non-interacting vortices and at intermediate fields it is described using activated flux flow. Eventually, the system attains a weakly localized metallic state at resistance value $R_{\rm C}$ as the magnetic field rises above a critical field $H_{\rm C}$. This critical point ($H_{\rm C},R_{\rm C}$) is seen to vary with $V_{\rm g}$. From the finite-size scaling analysis, this quantum-critical phase transition is found to lie in the (2+1)D XY universality class without disorder.

\section{Experimental Details}
\begin{figure}[h!]
	\centering
	\includegraphics[width=3.1in]{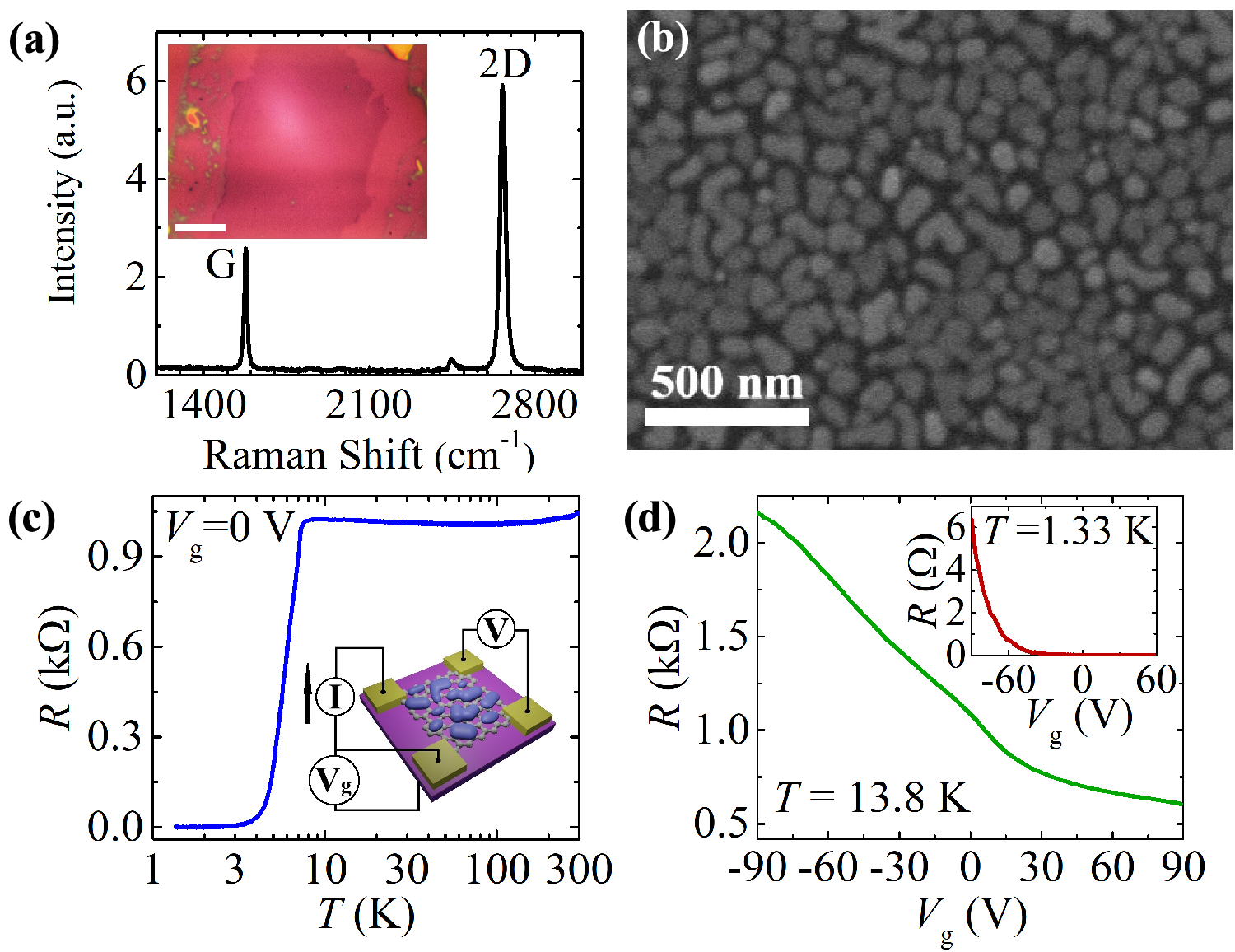}
	\caption{(a) Raman spectrum taken on graphene, shown in the inset, exhibiting the two peaks at approximately 1577 ${\mathrm{cm^{-1}}}$ (G peak) and 2665 ${\mathrm{cm^{-1}}}$ (2D peak) with intensity ratio I(2D)/I(G) = 2.3. The inset shows an optical image of this single-layer graphene with a 15 $\mathrm{\mu m}$ scale bar. (b) SEM image with Pb islands distributed on graphene. The dark grey regions correspond to Pb islands whereas the black background is graphene. (c) Temperature dependent resistance at $V_{\rm g}=0$ with the inset showing the four terminal electrical measurement schematic. (d) $V_{\rm g}$ dependence of the resistance at $T$ = 13.8 K showing that the Dirac point of the sample is at $V_{\rm g}<-90$ V. The inset shows the variation of resistance with $V_{\rm g}$ and at 1.33 K temperature.}
	\label{fig:Fig1}
\end{figure}

The samples were prepared by subjecting highly p-doped Silicon wafers with a 300 nm gate-quality oxide coating to 5 minutes each of sonication in acetone, isopropyl alcohol (IPA), and de-ionized water, followed by 2 minutes of cleaning in oxygen plasma at 50W \cite{huang2015reliable}. Kish graphite was then exfoliated onto these wafers within 30 minutes. After exfoliation, graphene flakes were identified using an optical microscope. The inset in figure \ref{fig:Fig1}(a) shows the optical image of the graphene layer of the studied device. The Raman spectrum of the graphene flake, shown in figure \ref{fig:Fig1}(a), confirmed it to be a single-layer graphene (SLG) as the ratio of the characteristic G and 2D bands is I(2D)/I(G) = 2.3. The absence of a D-peak indicates that the graphene is defect-free.

The electrical contacts on graphene were made by evaporating Cr/Au (5/45 nm) in a van der Pauw geometry using a mechanical mask to avoid contamination by chemicals and lithography resist. Subsequently, 30 nm of lead (Pb) was deposited on the graphene using the thermal evaporation technique at 20 \AA/sec rate, while maintaining the $\mathrm{SiO_2}$ substrate at 71$^{\circ}$C. The SEM image in figure  \ref{fig:Fig1}(b) shows that Pb formed discrete islands on graphene, with size ranging from 30 to 300 nm, instead of a uniform layer due to lead's poor wettability on graphene \cite{han2020disorder,han2020gate}. Lead deposited on the surrounding $\mathrm{SiO_2}$ substrate also formed distinct, well-separated islands, preventing electrical conduction through them. Therefore, electrical conduction occurs only through Pb islands coupled via the graphene.

After depositing Pb, the devices were immediately mounted on a cryostat, which was then cooled in a closed cycle refrigerator to its base temperature of 1.3 K. To minimize electromagnetic noise interference, low-pass R-C filters with a cutoff frequency of 15 kHz and pi-filters were installed in the measurement lines. Additionally, the measurement lines were routed through a low temperature Cu-powder filter in the sample holder to further reduce noise.

All transport measurements used a four-probe configuration with a DC current source as shown in the schematic in figure \ref{fig:Fig1}(c) inset. The resistance measurements were performed by biasing the device with 1 $\mu$A current of both polarities. The voltage from the device was amplified using a Femto Amplifier. A gate voltage ($V_{\rm g}$) ranging from $-$90 to 90 V was applied to the Si substrate with a 10 k$ \Omega$ series resistance. Magnetic field was applied perpendicular to the graphene plane by supplying electric current to a superconducting electromagnet. Temperature was measured using a Cernox temperature sensor placed close to the sample mounting plate in the sample holder. Magnetoresistive measurements of the sample were performed using a standard AC lock-in technique, with a bias current of 1 $\mu$A amplitude and 37 Hz frequency.

\begin{figure*}[h!]
	\centering
	\includegraphics[width=6.6in]{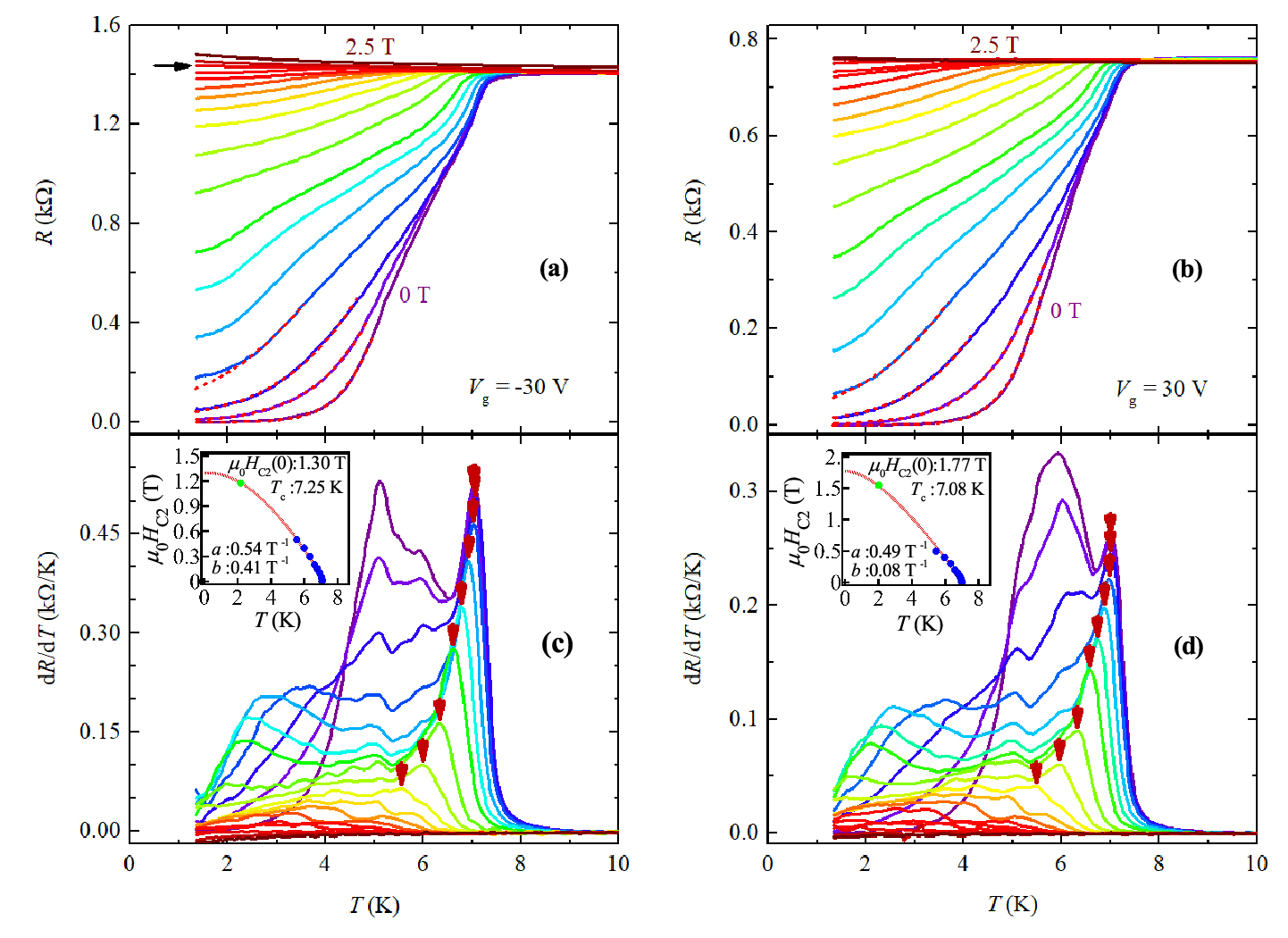}
	\caption{(a) and (b) Four probe resistance $R$ at 1 $\mu$A (bipolar) DC bias current as a function of temperature at $V_{\rm g} = -30$ and 30 V, respectively, for $\mu_0 H = $ 0, 0.007, 0.02, 0.05, 0.10, 0.15, 0.20, 0.30, 0.40, 0.50, 0.60, 0.70, 0.80, 0.90, 1.00, 1.20, 1.50 and 2.50 T. The red dashed lines are fits to equation (\ref{eq:Ambegaokar-Halperin}). (c) and (d) The derivative $\mathrm{d}R/\mathrm{d}T$ plots corresponding to (a) and (b), respectively, with the arrows marking the peaks corresponding to the onset transition temperature $T_{\rm CO}$. The insets in (c) and (d) show the corresponding variation of $\mu_0H_{\rm C_2}$ (blue dots) with temperature with the red dashed line showing fits to the WHH theory, equation (\ref{eq:WHH}), with $a, b$ and $T_c$ as the fitting parameters and the values for $\mu_0 H_{\rm C_2}(0)$ being deduced from the fits. The green dot corresponds to the critical field deduced from the scaling data discussed later.}
	\label{fig:Fig2}
\end{figure*}

\section{Results and discussion}
Figure \ref{fig:Fig1}(c) shows the variation of four-probe resistance from 300 K and down to 1.33 K base temperature for this hybrid Pb-graphene sample at $V_{\rm g} = 0$ V and at zero applied magnetic field. On the emergence of superconductivity inside the Pb islands, a drop in resistance occurs at a temperature around 7 K where the inter-island Josephson coupling sets-in and the resistance gradually goes to zero. A global phase coherent state appears with cooling as the Josephson coupling energy $E_{\rm J}$ rises. An increase in $V_{\rm g}$ from $-90$ to 90 V results in increment in electron density in graphene which also increases $E_{\rm J}$ \cite{allain2012electrical,han2014collapse}. A systematic variation of resistance with $V_{\rm g}$ at $T =$ 13.8 K in figure \ref{fig:Fig1}(d) shows that the Dirac point of this sample lies below $-90$ V, i.e. beyond the applied $V_{\rm g}$ range while figure \ref{fig:Fig1}(d) inset shows the $R(V_{\rm g})$ at $T =$ 1.33 K. Below we analyze the upper critical field as a function of temperature at different $V_{\rm g}$ values followed by the understanding of dissipation in this system with increasing field and eventually the transition to normal state that exhibits signatures of QPT.

\subsection{SC to normal transition and upper critical field}
Figures \ref{fig:Fig2}(a) and (b) display the temperature dependent resistance $R(T)$ curves at different field $\mu_0 H$ values from 0 to 2.5 T, for $V_{\rm g} = -30$ and 30 V, respectively, for the graphene-lead hybrid sample. The first derivative of the resistance with temperature, plotted in figure \ref{fig:Fig2}(c) and (d) for $V_{\rm g} = -30$ and 30 V, respectively, exhibits sharp peak at the superconductivity onset temperature $T_{\rm CO}$ which is close to 7 K for zero applied field. This $T_{\rm CO}$ is plotted as a function of $\mu_0 H$ in the insets of figures \ref{fig:Fig2}(c) and (d) showing decrement with increase in $\mu_0 H$. The red dashed line in these insets is the fit to Werthamer-Helfand-Hohenberg (WHH) model \cite{han2020disorder,werthamer1966temperature} which describes the boundary between the normal and superconducting regions in $T-\mu_0H$ plane,
\begin{align}
\ln\left(\frac{T_{\rm c}}{T}\right) =\Re\left[\Psi\left(\frac{1}{2} + \frac{(a+ib)\mu_0HT_{\rm c}}{2\pi T}\right)\right]-&\Psi\left(\frac{1}{2}\right).
\label{eq:WHH}
\end{align}
Here, $\Psi$ is the digamma function, $\Re$ refers to the real part and $T_{\rm c}$ is the zero field critical temperature. The Cooper pair breaking can occur in two different ways, namely: orbital pair breaking and spin or Pauli pair breaking. The parameters $a$ and $b$ determine the Maki parameter $\alpha=b/a$, which characterizes the relative strengths of the spin and orbital pair breaking \cite{maki1966effect}. The WHH model gives the temperature variation of upper critical field $H_{\rm C2}$ in the dirty limit where the coherence length is less than the mean free path.

\begin{figure*}
	\centering
	\includegraphics[width=6.7in]{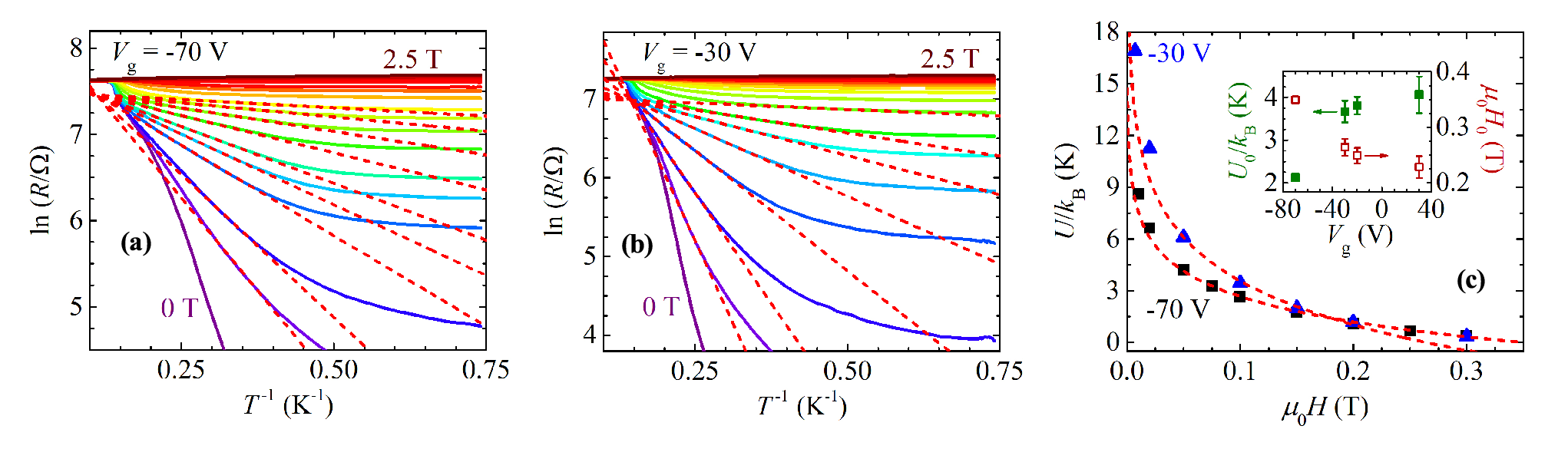}
	\caption{(a) and (b) Arrhenius plots of resistance $R$ at $V_{\rm g} = -70$ and $-30$ V, respectively, for different magnetic fields from 0 to 2.5 T. The red dashed lines in (a) are the fit to equation (\ref{eq:Arrhenius}) at $\mu_0 H = $ 0.01, 0.02, 0.05, 0.075, 0.10, 0.15, 0.20, 0.25, and 0.30 T. The red dashed lines in (b) are the fit to equation (\ref{eq:Arrhenius}) at $\mu_0 H = $ 0.007, 0.02, 0.05, 0.10, 0.15, 0.20, and 0.30 T. (c) Activation energies obtained from the slopes of the dashed lines as a function of magnetic field for $V_{\rm g} = -70$ (black squares) and $-30$ V (blue triangles). The red dashed lines are the fits to equation (\ref{eq:UH}). Inset shows plot of $U_0$ (closed squares) and $H_0$ (open squares) at different $V_{\rm g}$.}
	\label{fig:Fig3}
\end{figure*}

For our sample $\alpha<1$ and it reduces with increasing $V_{\rm g}$, see the insets of figures \ref{fig:Fig2}(c) and (d). This implies an increasing dominance of the orbital pair breaking with increasing $V_{\rm g}$. An earlier study on FeSe thin films \cite{stanley2024temperature} found that the Maki parameter reduces with reducing disorder. Thus, it can be inferred that the disorder in the lead-graphene system reduces with increasing $V_{\rm g}$. This can be expected as closer to the Dirac point the trap-induced charge puddles would lead to more disorder in graphene than away from it \cite{singh2018role,martin2008observation}. The fits also provide an estimate for the mean-field values of the upper critical field at zero temperature as listed in the insets of figures \ref{fig:Fig2}(c) and (d).

\subsection{Low field non-interacting vortex regime}
The red-dashed lines in figures \ref{fig:Fig2}(a) and (b) show the fits to the Ambegaokar-Halperin (AH) model that treats the vortex de-pinning as independent phase-slip processes \cite{suraina-PRB}. At low field the vortex density will be small and up to certain field these well-separated vortices can be assumed to be non-interacting. Further, a vortex crossing a Josephson junction of this JJ array is equivalent to a phase slip by $2\pi$ and such independent phase slip processes amount to a finite voltage and thus resistance. In this randomly distributed lead-islands' array the vortex pinning centers can arise at the meeting point of many islands as well as from the distribution of Josephson energy $E_{\rm J}$ values as further elaborated later. According to the AH model \cite{ambegaokar1969voltage}, the resistance of a single Josephson junction due to thermally activated phase slips is described by the following equation:
\begin{equation}
R = R_{\rm N}[I_{\rm 0}(\gamma/2)]^{-2}.
\label{eq:Ambegaokar-Halperin}
\end{equation}
Here, $I_0$ is the modified Bessel function of zero order and $\gamma$ is the normalized barrier potential given by expression $\gamma=A(1-T/T_{\rm CO})^m$, with $A$ and $m$ as magnetic field dependent constants \cite{suraina-PRB,bhalla2007vortex}. With increasing magnetic field vortex density increases and thus the resistance increases. Deviation from the fit to equation (\ref{eq:Ambegaokar-Halperin}) is observed for $\mu_0 H =$ 0.05 T for both the $V_{\rm g}$ values. This can arise because with increasing vortex density the inter-vortex interaction can not be ignored.

\subsection{Intermediate field thermally activated flux flow regime}
Figures \ref{fig:Fig3}(a) and (b) show $\ln R$ as a function of $1/T$, for $V_{\rm g} = -70$ and $-30$ V, respectively. The red dashed lines represent the Arrhenius equation that fits the data over a range of $1/T$ values for a given field. This equation is given by
\begin{equation}
R = R_0\, \exp\left(-\frac{U(H)}{k_{\rm B}T}\right)
\label{eq:Arrhenius}
\end{equation}
where, $R_0$ is a pre-factor, $k_{\rm B}$ is the Boltzmann constant, and $U(H)$ is the magnetic field dependent thermal activation energy. The above equation has been used to model thermally activated flux flow (TAFF) behavior of vortices in several superconducting systems \cite{han2020disorder,saito2015metallic,tsen2016nature}. Figure \ref{fig:Fig3}(c) shows the obtained $U$, plotted as a function of the magnetic field, for $V_{\rm g} = -70$ (black squares) and $-30$ V (blue triangles). The red dashed lines are the fits to the 2D thermally-assisted collective vortex-creep model, given by \cite{blatter1994vortices,feigel1990pinning},
\begin{equation}
U = U_0\: \ln\left(\frac{H_0}{H}\right).
\label{eq:UH}
\end{equation}
Here, $U_0$ is the pre-factor representing the vortex pinning potential \cite{han2020disorder} and $H_0$ is the magnetic field at which $U$ goes to zero, allowing free flow of vortices above this field \cite{saito2015metallic}. In this regime of flux flow with large vortex density the vortices' motion is also affected by the inter-vortex interactions \cite{feigel1990pinning} other than pinning. This leads to a collective motion of vortices which is well-supported by the fit to equation (\ref{eq:UH}) in figure \ref{fig:Fig3}(c).

\begin{figure*}
	\centering
	\includegraphics[width=6.6in]{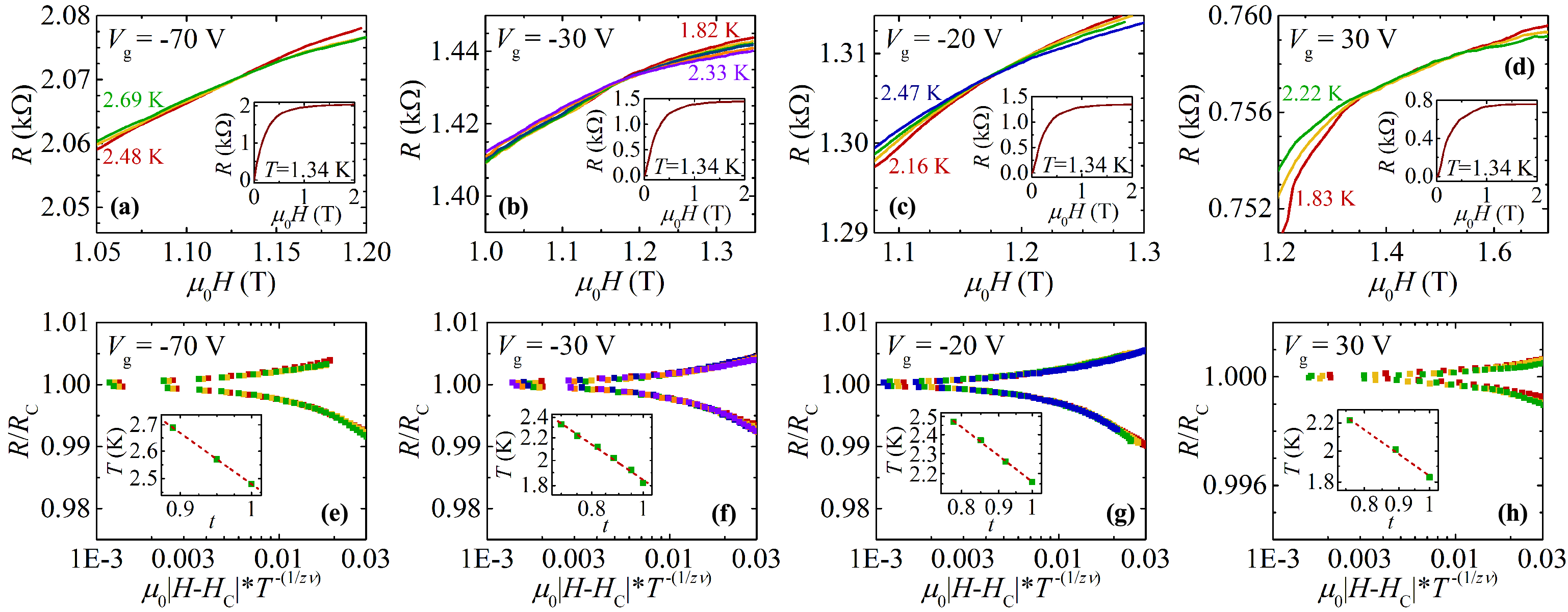}
	\caption{(a) Magnetoresistance $R(H)$ at temperatures, $T$ = 2.48, 2.57 and 2.69 K, for $V_{\rm g} = -70$ V. (b) $R(H)$ at $T$ = 1.82, 1.92, 2.02, 2.12, 2.22 and 2.33 K, for $V_{\rm g} = -30$ V. (c) $R(H)$ at $T$ = 2.16, 2.26, 2.37 and 2.47 K, for $V_{\rm g} = -20$ V. (d) $R(H)$ at $T$ = 1.83, 2.01 and 2.22 K, for $V_{\rm g} = 30$ V. Inset in these plots shows $R(H)$ at 1.34 K base temperature. Finite Size Scaling plot of $R/R_{\rm C}$ with $\mu_0|H-H_{\rm C}|T^{-1/z\nu}$ for $V_{\rm g} = -70$ (e), $-30$ (f), $-20$ (g) and 30 V (h). Inset in these plots shows $T$ Vs $t$ on log-log scale with fits to deduce $z\nu$.}
	\label{fig:Fig4}
\end{figure*}

It can be seen that the Arrhenius equation fits well in a temperature range slightly below $T_{\rm CO}$, where $E_{\rm J}$ is relatively small. Thus, the pinning barrier potential $U(H)$, proportional to $E_{\rm J}$ \cite{rzchowski1990vortex,lobb1983theoretical}, should also be relatively weak permitting the flux flow at finite temperatures. In this randomly distributed lead-islands' array there will be a distribution of $E_{\rm J}$ values and regions with lower than average $E_{\rm J}$ will favor vortices over those with higher $E_{\rm J}$. The voids formed at the intersection of several Pb islands will also act as pinning sites with higher $E_{\rm J}$ regions having higher depinning or activation energy. Apart from the structural inhomogeneity, the non-uniformity in the charge carrier density in graphene will lead to another form of disorder in the system. The interface defects between graphene and $\mathrm{SiO_2}$ result into the formation of charge puddles and this becomes more prominent for $V_{\rm g}$ close to the Dirac point of the graphene \cite{singh2018role,martin2008observation}. Thus an increase in disorder or a wider $E_{\rm J}$ distribution can be expected when the Dirac point is approached in this sample by reducing $V_{\rm g}$.

Figure \ref{fig:Fig3}(c) inset shows the plots of $U_0$ and $\mu_0 H_0$ for four different studied $V_{\rm g}$ values. With decreasing $V_{\rm g}$, $U_0$ is seen to decrease while $\mu_0 H_0$ shows an increasing trend. The former is expected as $E_{\rm J}$ reduces with decrement in $V_{\rm g}$ but the latter is somewhat surprising. This could arise from an interplay between the $V_{\rm g}$ dependence of $E_{\rm J}$ and disorder, or spread in $E_{\rm J}$. A reduction in $V_{\rm g}$ leads to an increase in disorder, as discussed above, but a decrease in $E_{\rm J}$. More disorder can increase the number of pinning sites while less $E_{\rm J}$ will reduce the depinning barrier. Eventually, the former could dominate over the later in dictating the $V_{\rm g}$ dependence of $\mu_0 H_0$.

With further increasing fields, the superconducting order parameter will reduce and diminish where the system will approach the normal state. At very low temperatures this transition to the normal metal state under an applied field will carry signatures of QPT which is discussed below for this Pb-graphene hybrid system.

\subsection{High field quantum critical regime}
With further increase in magnetic field, there is complete destruction of superconductivity in the lead islands and the system is driven towards a weakly localized metal. The $R(T)$ curve at $\mu_0 H =$ 1.2 T, marked by the black arrow in figure \ref{fig:Fig2}(a), represents the crossover between the superconducting region ($dR/dT > 0$) and the weakly localized metallic region ($dR/dT < 0$). Figure \ref{fig:Fig4}(a)-(d) show high resolution and low noise $R(H)$ measurements performed using lock-in technique at four different $V_{\rm g}$ values in narrow temperature ranges. For a fixed $V_{\rm g}$, the $R(H)$ curves in the narrow temperature range intersect at the same point $(\mu_0 H_{\rm C},R_{\rm C})$, defined as the critical point. Thus the resistance at this critical point is independent of temperature. For instance, figure \ref{fig:Fig4}(b) shows that several $R(H)$ plots for $V_{\rm g} = -30$ V for different temperatures intersect at the same point given by $\mu_0 H_{\rm C} =$ 1.177 T and $R_{\rm C} =$ 1.432 k$\Omega$. This is a signature of a continuous quantum phase transition. For $H < H_{\rm C}$, $dR/dT>0$, i.e. $R$ increases with $T$ while for $H > H_{\rm C}$, $dR/dT<0$. Note that $H_{\rm C}$ denotes the critical field associated with the QPT, which is close to $H_{\rm C2}$ as discussed later.

$R_{\rm C}$ in our case is significantly lower than the quantum resistance for Cooper pairs \cite{fisher1990quantum} $R_{\rm Q} = h/4e^2 = 6.45$ k$\Omega$. Yazdani et al. \cite{yazdani1995superconducting} observed a wide variation in the critical resistance values in $\mathrm{\alpha MoGe}$ thin films with values less than $R_{\rm Q}$. This apparent lack of universality was attributed to Fermionic excitations which lead to excess conductivity at the critical point, while the main characteristics of the quantum phase transition are still retained. A similar result with $R_{\rm C} < R_{\rm Q}$ has also been reported by Steiner et al. \cite{steiner2008approach} in weakly disordered $\mathrm{InO_x}$ film.

Finite size scaling (FSS) analysis as applied in previous experimental works \cite{sun2018double,xing2015quantum} is used here to understand the nature of the QPT. The FSS describes the resistance of the 2D system by a scaling law \cite{sondhi1997continuous,fisher1990quantum}
\begin{equation}
R = R_{\rm C} F(T^{-1/z\nu}\delta)
\label{eq:FSS}
\end{equation}
where, $\delta=\mu_0 |H-H_{\rm C}|$ is the absolute deviation from the critical magnetic field $\mu_0 H_{\rm C}$ and $F$ is an unknown function with $F$(0)=1. The parameter $\nu$ is correlation length exponent and $z$ is the dynamical scaling exponent. These are determined from the spatial correlation length $\xi$ and the temporal correlation length $\xi_\tau$, using the relations $\xi \sim \delta^{-\nu}$ and $\xi_\tau \sim \xi^z$ in the vicinity of the continuous QPT at $T = 0$ K  \cite{sondhi1997continuous}. These results are independent of the microscopic details of the transition but depend on system's dimensionality and the range of underlying interactions, which determine the universality class of the system.

The scaling exponent product $z\nu$ is found using a method \cite{sun2018double}, where each $R$ Vs $\mu_0 H$ curve at different temperatures is re-plotted in the form: $R/R_{\rm C}$ Vs $\mu_0 (H-H_{\rm C})$. The horizontal axis, i.e. $\mu_0 (H-H_{\rm C})$, of each curve is then scaled by a temperature dependent factor $t(T)$ so that all the curves collapse into the lowest temperature curve. The factor $t$ for the lowest temperature curve is chosen to be one as a convention. $T$ is then plotted as a function of $t$, after anticipating $t \propto T^{-1/z\nu}$, on a log-log scale, see the insets of figure \ref{fig:Fig4}(e)-(h). Finally, the slope of the linear fit to this plot gives the exponent $-z\nu$. As seen in figures \ref{fig:Fig4}(e)-(h), the $R(H)$ curves, when plotted with respect to the scaling variable $\mu_0|H-H_{\rm C}|T^{-1/z\nu}$, collapse onto a single curve representing the unknown function $F(\mu_0|H-H_{\rm C}|T^{-1/z\nu})$. This collapse confirms the existence of the quantum critical behavior. The dynamical exponent $z$ can be assumed to be one which corresponds to the long range Coulomb interaction between charges \cite{fisher1990quantum,fisher1990presence,yazdani1995superconducting}.

Table \ref{tab:table1} lists the values of $\mu_0 H_{\rm C}$, $R_{\rm C}$ and the critical exponent $\nu$ for different studied $V_{\rm g}$ values. There is a monotonic rise in $\mu_0 H_{\rm C}$ with increase in $V_{\rm g}$ along with a decrease in $R_{\rm C}$ while the critical exponent $\nu$ for all the studied $V_{\rm g}$ values is close to 2/3. The slight variation observed in the value of $\nu$ can be attributed to limited range of temperature and magnetic field used for the scaling analysis. An exponent $\nu$ being close to 2/3 is consistent with the universality class corresponding to the $(2+1)$D XY model without disorder \cite{sondhi1997continuous,cha1994universal}.

The critical exponent $\nu \sim 2/3$ was also reported for the magnetic field-tuned SIT in conventional 2D thin film systems, such as $a$-Bi \cite{markovic1999superconductor}, $a$-NbSi \cite{aubin2006magnetic}, $a$-WSi \cite{zhang2018sequential} and underdoped $\mathrm{La_{2-x}Sr_xCuO_4}$ \cite{shi2014two}. In these systems, the studied critical regions lie close to $T_{\rm CO}(0)$ where the Cooper pairing is nearly destroyed and thus the critical field in these systems is characterized as the depairing field of the Cooper pairs. Magnetic field-tuned SIT observed in granular 2D systems such as $\mathrm{LaTiO_3/SrTiO_3}$ interface \cite{biscaras2013multiple}, $\mathrm{LaAlO_3/KTaO_3}$ interface \cite{chen2021electric} and tin-graphene hybrid system \cite{sun2018double} show similar critical exponent for the critical magnetic field.

For the studied Pb-graphene hybrid sample, the critical field $H_{\rm C}$ for QPT is located on the WHH curve, see the green dot in the insets of figure \ref{fig:Fig2}(c,d). Thus this $H_{\rm C}$ is same as or close to the $H_{\rm C2}$ where the order parameter diminishes due to pair breaking. Thus, the pair-breaking effects are likely to dominate the studied QPT due to the diminishing amplitude of the superconducting order parameter \cite{hsu1993magnetic} inside the Pb islands. While it is understood that the zero field SIT, as a function of disorder or the ratio $E_{\rm C}/E_{\rm J}$, is bosonic as the SC order parameter remains intact throughout. On the other hand, in the transition from a SC to a bad metal state, driven by the magnetic field, the SC order parameter diminishes at the transition and thus Fermionic physics will play a dominant role. However, one cannot completely rule out the bosonic physics as some superconducting correlations, however small, will still persist \cite{skocpol1975fluctuations} slightly above this field.
\begin{table}[h!]
  \begin{center}
    \caption{The values of critical magnetic field $\mu_0 H_{\rm C}$, critical resistance $R_{\rm C}$ and critical exponent $\nu$ for different gate voltage $V_{\rm g}$.}
    \label{tab:table1}
    \begin{tabularx}{0.48\textwidth} {
  | >{\centering\arraybackslash}X
  | >{\centering\arraybackslash}X
  | >{\centering\arraybackslash}X
  | >{\centering\arraybackslash}X | }
     \hline
      \text{$V_{\rm g}$ (V)} & \text{$\mu_0 H_{\rm C}$ (T)} & \text{$R_{\rm C}\: (k\Omega)$} & \text{$\nu$}\\
      \hline
      -70 & 1.127 & 2.070 & 0.70\\
      -30 & 1.177 & 1.432 & 0.67\\
      -20 & 1.176 & 1.307 & 0.54\\
       30 & 1.547 & 0.758 & 0.70\\
     \hline
    \end{tabularx}
  \end{center}
\end{table}

\section{Summary and conclusions}
The gate voltage dependent transition to normal state from superconducting state under perpendicular magnetic field in a Josephson junction array comprising of randomly distributed Pb islands on exfoliated single-layer graphene shows three interesting transport regimes. At low fields the dynamics of non-interacting vortices is dominated by activated de-pinning processes which is followed by collective flux flow of interacting vortices at intermediate fields with a field dependent activation barrier. The field dependence of the barrier is dictated by the gate voltage as the inter-island Josephson coupling increases with the gate voltage while the disorder reduces. A quantum critical regime is found near the upper critical field where the superconducting order parameter in the lead islands diminishes. From the finite size scaling analysis the critical exponent $\nu$ is found to be close to 2/3, which corresponds to the universality class of $(2+1)$D XY model.

\section*{Acknowledgements}
We thank Pratap Raychaudhuri for discussions on device details and measurements. We acknowledge SERB-DST of the Government of India and IIT Kanpur for financial support.

\end{document}